\documentclass[a4paper,12pt]{article}
\usepackage{amsmath}
\usepackage{verbatim}
\usepackage{amssymb}
\usepackage{mathrsfs}
\usepackage{inputenc}
\usepackage{textcomp}
\usepackage{appendix}

\newcommand{\be}[1]{\begin{equation}\label{#1} }
\newcommand{\ee}{\end{equation}}
\newcommand{\bea}[1]{\begin{eqnarray}\label{#1} }
\newcommand{\eea}{\end{eqnarray}}

\def\lp{\ell_P}
\begin{document}

\title{ \vspace{-0.5cm}
\bf Entanglement Entropy for $T{T}$ deformed CFT in general dimensions}
\date{}

\author{\!\!\!\!Aritra Banerjee$^{x}$\footnote{aritra@itp.ac.cn}, Arpan Bhattacharyya$^{y}$\footnote{bhattacharyya.arpan@yahoo.com} and Soumangsu Chakraborty$^{z}$\footnote{soumangsuchakraborty@gmail.com}\\ ~~~~\\
\it ${^x}$ CAS Key Laboratory of Theoretical Physics, Institute of Theoretical Physics, \\Chinese Academy of Sciences,  Beijing 100190, P.R. China.\\
\it ${^y}$Center for Gravitational Physics, Yukawa Institute for Theoretical Physics,\\
Kyoto University, Kitashirakawa Oiwakecho, Sakyo-ku, Kyoto 606-8502, Japan. \\ 
\it${^z}$Racah Institute of Physics,
The Hebrew University, Jerusalem 91904, Israel.}
\maketitle

\vskip 2cm

\abstract{ We consider deformation of a generic $d$ dimensional ($d\geq 2$) large-$N$ CFT on a sphere by a spin-0 operator which is bilinear in the components of the stress tensor. Such a deformation has been proposed to be holographically dual to an $AdS_{d+1}$ bulk with a hard radial cut-off. We compute the exact partition function and find the entanglement entropy from the field theory side in various dimensions and compare with the corresponding holographic results.  We also compute renormalized  entanglement entropy both in field theory and holography and find complete agreement between them.}

\newpage

\tableofcontents

\section{Introduction}

In the past couple of years there has been a growing interest in deformations of two dimensional quantum field theories (QFT) by a class of irrelevant operators which are bilinear in conserved currents \cite{Smirnov:2016lqw}. Such deformations of a 2d QFT turn out to be solvable and if the undeformed theory to start with is integrable, then the deformations also preserve integrability. The deformed theory turns out to be perfectly well defined in spite of the fact that these deformations involve flowing up the renormalization group (RG). Among the  known examples of such deformations, aspects of the so-called $T\bar{T}$ \cite{Smirnov:2016lqw,Cavaglia:2016oda} deformation (where  $T$ and $\bar{T}$ are respectively the holomorphic and anti-holomorphic components of the stress tensor in two dimensions) are particularly well studied \cite{McGough:2016lol,Shyam:2017znq,Giveon:2017nie,Giveon:2017myj,Asrat:2017tzd,Chakraborty:2018kpr,Chakraborty:2018aji,Dubovsky:2017cnj,Dubovsky:2018bmo,Giribet:2017imm,Kraus:2018xrn,Cardy:2018sdv,Cottrell:2018skz,Aharony:2018vux,Cardy:2018jho,Dubovsky:2018dlk,Bonelli:2018kik,Datta:2018thy,Donnelly:2018bef,Babaro:2018cmq,Conti:2018jho,Chen:2018eqk,Aharony:2018bad,Jiang:2019tcq,Park:2018snf,Sun:2019ijq,Wang:2018jva,Gorbenko:2018oov,Araujo:2018rho}.  
Deformation of a 2d conformal field theory (CFT$_2$) in the infrared (IR) by $T\bar{T}$ breaks conformal invariance but preserves Lorentz invariance. The theory in the ultraviolet (UV) is non-local in the sense that the high energy behavior is not governed by a fixed point. A realization of a $T\bar{T}$ deformed QFT$_2$ as a theory of quantum gravity is discussed in \cite{Dubovsky:2017cnj,Dubovsky:2018bmo}.

The other known examples of solvable irrelevant deformations include $J\bar{T}$ (where $J$ is a conserved holomorphic $U(1)$ current) and some general linear combination of  $T\bar{T}$ and $J\bar{T}$ \cite{Guica:2017lia,Bzowski:2018pcy,Chakraborty:2018vja,Aharony:2018ics,Guica:2019vnb,Giveon:2019fgr,LeFloch:2019rut, Apolo:2018qpq, Nakayama:2018ujt}. In both these cases the deformation breaks conformal as well as Lorentz invariance.

The spectrum of a $T\bar{T}$ deformed 2d CFT turns out to be sensitive to the sign of the $T\bar{T}$ coupling. For one sign of the coupling, the so-called good sign, the spectrum of the deformed theory is real and the theory is unitary. At high energies, the theory exhibits Hagedorn density of states much like 2d Little String Theory \cite{Giveon:2017nie}.  However for the other sign of the coupling, the so-called wrong sign, the energies of the highly excited states become complex and the theory turns out to be non-unitary. For the wrong sign of the $T\bar{T}$ coupling, there has been a proposal for a holographic dual in the form of $AdS_3$ with a sharp cut-off \cite{McGough:2016lol}. This particular concept was further developed in \cite{Shyam:2017znq,Kraus:2018xrn,Cottrell:2018skz,Donnelly:2018bef}. A different holographic interpretation of a ``single-trace" $T\bar{T}$ deformed CFT$_2$ (considering the theory on a  symmetric product orbifold)  appears in   \cite{Giveon:2017nie,Giveon:2017myj,Asrat:2017tzd,Chakraborty:2018kpr,Chakraborty:2018aji,Giribet:2017imm} for both sign of $T\bar{T}$ coupling. The holographic interpretation of $J\bar{T}$ deformed CFT$_2$ appears in \cite{Bzowski:2018pcy,Chakraborty:2018vja}.

Inspired by the two dimensional deformed setup, generalization of $T\bar{T}$ deformation to higher dimension has been proposed in \cite{Taylor:2018xcy,Hartman:2018tkw} and further developed in \cite{Caputa:2019pam}. In an attempt to understand holography with a hard radial cut-off in $AdS_{d+1}$ with Dirichlet boundary condition on the cut-off surface, the authors of \cite{Taylor:2018xcy,Hartman:2018tkw} constructed a dual effective field theory. This effective theory at long distances appears to be a large-$N$ CFT$_d$ deformed by an irrelevant operator, $X_d$, which is bilinear in the components of the stress tensor and transforms as a spin-0 operator under Lorentz transformation. This operator exactly reduces to the $T\bar{T}$ operator in two dimensions and can be thought of as a higher dimensional generalization of the $T\bar{T}$ operator. 

The construction of the deformation operator $X_d$ in arbitrary dimensions on an arbitrary background have been elucidated in \cite{Hartman:2018tkw,Caputa:2019pam}. Of course the operator $X_d$ for $d\geq 3$, in general, does not posses all the nice properties as the  $T\bar{T}$ operator in two dimensions (see \cite{Zamolodchikov:2004ce} for details). In particular, there is no reason to believe that it will satisfy the well known factorization property of the  
$T\bar{T}$ operator in two dimensions in a non-compact flat spacetime. However, restricting ourselves to the regime of large-$N$ field theories, one can exploit large-$N$ factorization property of multi-trace operators and study the deformation of a generic large-$N$ CFT$_d$ at strong coupling by the operator $X_d$.  

In this paper, we study the deformation of a large-$N$ CFT$_d$ on a sphere $S^d$ deformed by the operator $X_d$. We work in regime where  $N$ is large and $\lambda_d$ (where $\lambda_d$ is coupling to the operator $X_d$) is small\footnote{For finite $\lambda_d$,  one needs non-perturbative (in $\lambda_d$) realization of the deformation which, so far, is not well understood.} such that $N\lambda_d^{2/d}$ is positive\footnote{For negative values of the coupling $\lambda_d$, the asymptotic density of states at high energies is super-Hagedorn (density of states $\sim \exp\left(E^{\frac{2(d-1)}{d}}\right)$ where $E$ is the energy). Surprisingly, the scaling matches with the density of states of $D-(d-1)$ branes in the semiclassical approximation  \cite{Duff:1987cs}. For $d=2$, this is related to the Hagedorn spectrum of LST \cite{Giveon:2017nie,Giveon:2017myj,Asrat:2017tzd,Chakraborty:2018kpr,Chakraborty:2018aji}.} and finite. We compute the non-perturbative (in $N\lambda_d^{2/d}$) partition function from the field theory side and show that there is an exact match with the one derived from holographic analysis \cite{Caputa:2019pam}. We calculate the exact entanglement entropy between two co-dimension one hemispheres. Here, we find the entanglement entropy to be finite and independent of the UV cut-off. It's not surprising that the entanglement we calculate is free of UV divergence. In fact the field theory technique that we have adapted in this paper eliminates the non-universal polynomial terms in powers of $R$ (radius of the sphere), in particular the \textit{`area law'} term. However, the universal terms namely the logarithmic and constant terms are always retained in the process.  


A natural question that arises at this point is: \textit{does the entanglement entropy calculated from the field theory side agree with the entanglement entropy computed using Ryu-Takayanagi (RT) formula 
\cite{Ryu,Ryu1}  in $AdS_{d+1}$ with a hard radial cut-off ?} The RT minimal surface picks up both the universal as well as the non-universal terms in the entanglement entropy. Of course the entanglement entropy calculated from the field theory side matches perfectly with the universal terms of the holographic entanglement entropy \cite{Ryu,Ryu1,SM, SM1, Solo1, Solo2} \footnote{In  2d, an exact match between the result from field theory and holography is demonstrated in \cite{Donnelly:2018bef}.}. Moreover, switching off all the counterterms in the field theory side (i.e. working with the bare partition function) one could obtain the entanglement entropy which matches exactly with the holographic computation. A more appropriate comparison would be to first define the `Renormalized Entanglement Entropy' (REE)  following \cite{LiuMZ, LiuMZ1} for both field theory and holography \footnote{For related connection between REE and renormalized volume, one can see \cite{renvol} and related references.}. As shown in \cite{LiuMZ, LiuMZ1} area terms (both leading and sub-leading ones) are being naturally excluded from the structure of REE. We compute REE for our case both from field theory and holography and find a perfect agreement between them in various dimensions. Curiously, we also observe that REE interpolates between a zero value at high energies ($R\rightarrow 0,$ $R$ is the radius of the entangling surface) and the CFT central charge at the IR  fixed point ($R \rightarrow \infty$) for even dimensions \cite{Ryu,Ryu1,SM}. For odd dimensions it coincides with finite part of sphere partition function (independent of $R$) at IR fixed point. This nicely matches with our intuition based on already available results in the literature for two dimensions. 

The organization of this paper is the following. In section (\ref{sec2}) we discuss $T T$ deformation in general dimensions and introduce necessary ingredients to compute the sphere partition function. In section (\ref{sec3}) we compute the entanglement entropy between two codimension-one hemispheres from  field theory viewpoint. In section (\ref{sec4}) we compute the entanglement entropy using RT formula in a $AdS$ geometry with a finite radial cut-off and compare  our results with the field theory ones.  In section (\ref{sec5}) we move on to compute REE and find perfect agreement between  both field theory and  holographic computations. Lastly, we conclude our paper with a discussion mentioning various future directions. Some additional details about exact matching of EE from both sides are kept into the appendix.

\section{$T{T}$ deformation in general dimensions: Sphere partition functions}\label{sec2}
In what follows, we will be computing the sphere partition function in a $TT$\,\footnote{From now on we will call  the bilinear operator $X_d$ as the $TT$ operator or $X_d$ interchangeably.} deformed CFT living on a sphere in $d\geq 2$ dimensions. On our way, we will consider necessary ingredients to evaluate the partition function and introduce methods to to extract the exact entanglement entropy. We will also give a very brief review of constructing the deformation operator $X_d$ following  \cite{Taylor:2018xcy,Hartman:2018tkw}.  In a later section, we will be presenting the explicit results in various dimensions.
\subsection{General outline}
Let us consider a $d$ dimensional generic Lorentz invariant QFT living on a sphere of radius $R$. We would like to calculate the exact partition function, $Z_{S^d}$, of such a general setup. Let us consider the metric, $\gamma_{ab}$ on the sphere, to be of the following form: 
   \begin{eqnarray}
ds^2&=& R^2\left(d \theta_1^2+\sum_{j=2}^{d} \prod_{i=1}^{j-1}\sin^2 \theta_i d \theta_j^2\right)
\end{eqnarray}
where $\theta_i\in (0,\pi)$ for $i=1,2\hdots ,d-1$ and $\theta_d\in(0,2\pi)$.
The change of the partition function due to an infinitesimal deformation of the metric is given by,
\begin{eqnarray}\label{varpf}
\delta \log Z_{S^d}=-\frac{1}{2}\int d^dx\sqrt{\gamma}\langle T^{ab}\rangle\delta \gamma_{ab}.
\end{eqnarray}
where $T_{ab}$ are the components of the stress tensor of QFT. Equation \eqref{varpf} holds for any variation of the background metric. Thus the response of the partition function due to a variation in the radius of the sphere is given by
\begin{eqnarray} \label{part}
 R\frac{\partial}{\partial R}\log Z_{S^d}=-\int d^d x \sqrt{\gamma}\langle T^a_a\rangle.
 \end{eqnarray}

  Since we are interested in computing the entanglement entropy between two codimension-one hemispheres, we introduce below the metric on the replicated sphere (also known in the literature as the $n$-sheeted sphere):
 \begin{eqnarray}
ds^2&=& R^2\left(d \theta_1^2+\sum_{j=2}^{d-1} \prod_{i=1}^{j-1}\sin^2 \theta_i d \theta_j^2+n^2\prod_{i=1}^{d-1}\sin^2 \theta_i d \theta_d^2\right),
\end{eqnarray}
where $\theta_i\in (0,\pi)$ for $i=1,2\hdots ,d-1$, $\theta_d\in(0,2\pi)$ and $n$ is a positive integer. In our notation the coordinate $\theta_d$ is normal to the  two hemispherical caps whose entanglement we want to compute.  We denote the metric on the replicated sphere by $\gamma_{ab}^n$ and the replicated partition function by $Z_{S^d}^n$. Note that \eqref{varpf} holds for any QFT on an arbitrary manifold. Thus the variation of the partition function on the replicated sphere is given by
\begin{eqnarray}\label{varpfr}
\delta \log Z^n_{S^d}=-\frac{1}{2}\int d^dx\sqrt{\gamma^n}\langle T^{ab}\rangle_n \delta \gamma^n_{ab},
\end{eqnarray}
where $\langle \cdots \rangle_n$ denotes the expectation on the replicated sphere. The response of the partition function due to change in the replica index around $n=1$ is given by
\begin{eqnarray}\label{partn}
\frac{\partial}{\partial n}\log Z^n_{S^d}\Big{|}_{n=1}=-\frac{1}{d}\int d^dx \sqrt{\gamma} \langle T^a_a\rangle.
\end{eqnarray}
To get to the right hand side of the above equation we have used the symmetries on a sphere which dictate,
\begin{eqnarray}
\langle T_{ab}\rangle &=&\omega_d(R) \gamma_{ab},\label{symT}\\
\langle T^{\theta_d}_{\theta_d}\rangle &=&\frac{1}{d}\langle T^a_a\rangle,
\end{eqnarray}
where the proportionality function $\omega(R)$ depends on the the radius of the sphere ($R$), and this function needs to be determined.\par

Substituting \eqref{symT} in \eqref{part}, one obtains
\begin{eqnarray} \label{eq2.9}
R\frac{\partial}{\partial R}\log Z_{S^d}=-d\int d^d x \sqrt{\gamma}\omega_d(R).
\end{eqnarray}

The entanglement entropy, $S_{d,EE}$ is obtained by analytically continuing in $n$ and taking the limit $n\to 1$:
\begin{eqnarray}\label{EEdef}
S_{d, EE} &=&\left(1-n\frac{d}{dn}\right)\log Z^n_{S^d}\Big{|}_{n\to 1}.
\end{eqnarray}
Thus from \eqref{part}, \eqref{partn} and \eqref{EEdef} one can conclude that
\begin{eqnarray}\label{SEE}
S_{d, EE} &=&\left(1-\frac{R}{d}\frac{d}{dR}\right)\log Z_{S^d}.
\end{eqnarray}    
    
   Thus, this procedure gives us a simple but elegant way of extracting the entropy from the sphere partition function.
  
   \subsection{The deforming operator $X_d$} 
   
  As discussed in the introduction, the main motivation in constructing the higher dimensional generalization of the $T\bar{T}$ operator comes from understanding  holography in $AdS_{d+1}$ with a hard radial cut-off \cite{Taylor:2018xcy,Hartman:2018tkw}. As it turns out, for the holographic duality with cut-off $AdS$ to make sense, the dual effective field theory (EFT) must be deformed by an operator $X_d$. The operator $X_d$ turns out out to be bilinear in the components of the stress tensor, has spin-0 and has mass dimension $2d$. We restrict ourselves in the limit of large-$N$ and small $\lambda_d$ (with $\lambda_d\geq 0$) such that the quantity $N\lambda_d^{2/d}$ is finite. In this limit the operator $X_d$ takes the form,
 \begin{eqnarray} \label{eq1}
 X_d=-\frac{1}{d\lambda_d}  T^a_a.
 \end{eqnarray}

 For the sake of completeness, we present below the structure of the operators $X_d$ in general dimensions. In this paper we closely follow the conventions used in \cite{Hartman:2018tkw, Caputa:2019pam} and the reader is directed to these works for a more comprehensive understanding. 
   
 Using the holographic (Brown-York) stress-tensors  and deformation of the classical bulk gravity action in $AdS_{d+1}$ with hard radial cut-off, the authors of \cite{Hartman:2018tkw} have derived the EFT deforming operator $X_d$ given by \footnote{ Here for any tensor $M_{ab}$, $M_{ab}^2= M_{ab}M^{ab}$ where index are are raised by the metric on sphere.}
\begin{align}\label{Xdef}
X_d =  \left(T_{ab}+\frac{\alpha_d}{\lambda_d^{\frac{d-2}{d}}}\mathcal{C}_{ab}\right)^2-\frac{1}{d-1}\left(T^{a}_a+\frac{\alpha_d}{\lambda_d^{\frac{d-2}{d}}}\mathcal{C}^a_a\right)^2+\frac{1}{d }\frac{\alpha_d}{\lambda_d^{\frac{2(d-1)}{d}}}\left(\frac{(d-2)}{2}\mathcal{R}+\mathcal{C}^a_a\right).
\end{align}
Here  $\alpha_d$ is a dimensionless parameter which characterizes the degrees of freedom of the undeformed CFT.    
Evidently when $d=2,$ one recovers the results of \cite{Donnelly:2018bef}. One should also note that the tensor $\mathcal{C}_{ab}$ is non vanishing only for $d\geq 3.$  In this paper, we would explicitly consider the cases $d=2,3,4,5,6.$ For being complete, let us write the explicit expressions  for $\mathcal{C}_{ab}$ below for various dimensions following \cite{Hartman:2018tkw,Caputa:2019pam} \footnote{For  the expression for $\mathcal{C}_{d}$ for general dimensions please refer to the appendix (A).}, 
\begin{align}
\begin{split}
&d =3, 4 : \mathcal{C}_{ab}=\mathcal{G}_{ab},\\& 
d=5,6 : \mathcal{C}_{ab}= \mathcal{G}_{ab}+\frac{2\, d \, \alpha_d \,\lambda_d^{\frac{2}{d}}}{d-4}\Big[2(\mathcal{R}_{acbd}\mathcal{R}^{cd}-\frac{1}{4}\gamma_{ab} \mathcal{R}_{cd}\mathcal{R}^{cd})-\frac{d}{2(d-1)}(\mathcal{R} \mathcal{R}_{ab}-\frac{1}{4}\gamma_{ab}\mathcal{R}^2)\Big].
\end{split}
\end{align}
Here $\mathcal{G}_{ab}$ is the Einstein tensor and $\mathcal{R}_{abcd}$ are the Riemann curvature of the background manifold where the field theory lives. Note that the the operator $X_d$ in \eqref{Xdef} results from replacing a particular bulk observable by their boundary counterpart. The end product gives an  EFT operator that can be viewed as an obvious generalization of $T\bar{T}$ operator to higher dimensions.  For the deformed EFT of our interest, which resides on a $d$ dimensional sphere of radius $R$, one can solve $\omega_d(R)$ defined in \eqref{symT} from the following equation 
\begin{eqnarray} \label{eq2}
 \langle T^a_a\rangle = - d\, \lambda_d \langle X_d\rangle.
\end{eqnarray}

\subsection{On Holographic dictionary}
As we will compare our field theoretic result for $S_{d, EE}$ with the results from holography, we give here a simple sketch of the dictionary .  As mentioned earlier the holographic dual to this deformed theory in $d$ dimensions is a $AdS_{d+1}$ spacetime with a finite cut-off along the radial direction. The metric of $AdS_{d+1}$ is shown below,
\be{} \label{bulkm}
ds^2=L^2 \Big(\frac{dr^2}{r^2}+ r^2 \gamma_{ab} dx^a dx^b\Big),
\ee
where $L$ is the AdS radius and $\gamma_{ab}$ is a $d$ dimensional Euclidean metric on the cut-off surface. For the computation of the entanglement entropy, it is always better to work in  Euclidean  signature. The  hard radial cut-off is placed at $r= r_c.$ We consider the bulk gravity theory to be  described by the Einstein-Hilbert action. The bulk action with suitable surface and counter terms is shown below,
\be{} \label{bulkaction}
S_{tot}=S_{bulk}+S_{surf}+S_{ct}\\,
\ee
\be{}
S_{bulk}=-\frac{1}{2 \lp^{d-1}}\int d^{d+1}x\sqrt{g}(\hat R+\frac{d(d-1)}{L^2})\,,\\
\ee

\be{}
S_{surf}=-\frac{1}{\lp^{d-1}}\int d^{d}x\sqrt{h} ~\mathcal{K}\,,\\\ee
where $\hat R$ is the bulk curvature.  $\mathcal{K}_{ab}= h^{c}_{a}h^{d}_{b}\nabla_{c}\hat n_{d}$  is the extrinsic curvature defined on the cut-off surface $r=r_c$ with the induced metric $h_{ab}=g_{ab}-\hat n_{a}\hat n_{b}$. Also in our notation, $g_{ab}$ is the bulk metric shown in (\ref{bulkm}), $\hat n_{a}$ is the unit normal  and $\mathcal{K}= h^{ab}\nabla_{a}\hat n_{b}$ is trace of the extrinsic curvature. Note that we have written all normalizations in the form of the Planck length ($\lp$) to maintain uniformity throughout the manuscript. For convenience, we mention that $\lp^{d-1} = 8\pi G_N$, where $G_N$ is the Newton constant in $d+1$ dimensional bulk.

The counterterm action on the other hand is given by \cite{Emparan:1999pm,me},
\be{} \label{ict}
S_{ct}=\frac{1}{\lp^{d-1}}\int d^{d}x \sqrt{h} \left[c_1\, \frac{d-1}{ L} + \frac{c_{2}~L}{2(d-2)} \mathcal{R}+\frac{ c_{3}~L^{3}}{2(d-4)(d-2)^{2}}(\mathcal{R}_{ab}\mathcal{R}^{ab}-\frac{d}{4(d-1)}\mathcal{R}^{2})+...\right]\,,\\
\ee
where $\mathcal{R}_{ab}$ is the Ricci curvature tensors of the cut-off surface and we define $\mathcal{R}=\mathcal{R}_{ab}h^{ab}.$ The constants used in the action, $c_1=1$ and is non-vanishing only  for $d \geq  2$, $c_2=1$ and is non-vanishing only  for $d \geq 3$ and lastly, $c_3 =1$ and is non-vanishing only for $d\geq 5$ and so on. For a more detailed exposition to the structure of these counter terms, the reader could refer to \cite{Emparan:1999pm, me, Hartman:2018tkw}. \footnote{The structure of counterterms are so chosen that it cancels the UV divergent pieces in the partition function, see Appendix A for details.}

 Given the action (\ref{bulkaction}) one can now compute the bulk Brown-York stress tensor ($T^{BY}_{ab}$) following \cite{Balasubramanian:1999re}. The field theory stress tensor (on the Boundary) is related to the bulk one in the following way,
\be{}
T_{ab} = r_c^{d-2} T^{BY}_{ab}.
\ee
For simplicity henceforth we will set $r_c=1$ throughout the calculation. Also to establish the dictionary, one can express all the field theory parameters in terms of parameters of gravitational theory. This gives rise to,
\begin{align}
\begin{split} \label{relation}
\lambda_d=\frac{\lp^{d-1} L}{2\,d},\,\,\, \alpha_d=\frac{L^{\frac{2(d-1)}{d}}}{(2 d)^{\frac{d-2}{d}}(d-2)\lp^{\frac{2(d-1)}{d}}},\,\,\, L^2= 2d(d-2)\alpha_d\lambda_d^{2/d}.
\end{split}
\end{align} 
We will use these relations frequently to make comparison with holographic results. \par 

Let us illustrate briefly the machinery we have built upto now, notice that using the recipe mentioned before for the construction of $X_d$ and the dictionary (\ref{relation}),  $X_2$ takes  the following form ($\mathcal{C}_{ab}=0$ for $d=2$),
\be{}\label{X2}
X_2=T_{ab}T^{ab}-(T^{a}_a)^2+\frac{L}{4\,\lambda_2 \lp }\mathcal{R}.
\ee
Here we denote the two dimensional  central charge by $c,$ related to bulk the quantities  via holographic dictionary as \cite{Central},
\be{} \label{2c}
c=\frac{12 \pi L}{\lp}.
\ee 
Using  this we recover the result of \cite{Donnelly:2018bef} for 2-dimensions, complete with the usual anomaly term,
\be {}
X_2 =T_{ab}T^{ab}-(T^{a}_{a})^2+\frac{1}{2\lambda_2}\,\frac{c}{24\pi} \mathcal{R}.
\ee
\subsection*{Expansion parameter:} 
As mentioned in the beginning  we are studying the this deformation for a generic large $N$ CFT.  We are working in the limit where $N$ is very large, $\lambda_d^{2/d}$ is small but $N \lambda_d^{2/d}$ is finite. This enables us to define the following expansion parameter. For $d>2$  we have,
\be{}\label{expand}
t_d=\alpha_{d}\,\lambda_d^{2/d}.
\ee
Let us remind that $\alpha_d$ counts the number of degrees of freedom in the undeformed CFT  and is evidently proportional to $N$, the rank of the gauge group\,\footnote{For example in $\mathcal{N}=4$ SYM with $SU(N)$ gauge group in $4d$ we have $\alpha_4=\frac{N}{2^{7/2}\pi}$. In $d=3$ for ABJM theory, $\alpha_3=\frac{N}{6 (2\pi^2)^{1/3}}$ where $N$ is the rank of the gauge group in ABJM \cite{Caputa:2019pam}.}.  Although our result will be valid for any values of $t_d,$ nonetheless to make comparison with known results of undeformed CFT we will frequently expand our result in the limit $R \gg \sqrt{t_d}.$

For two dimensions, the corresponding expanding parameter will be,
\be {}
t_2= c\,\lambda_2.
\ee
Again we can expand our results in $2d$ in the limit $R \gg \sqrt{t_2}$ to recover the results of CFT \cite{Donnelly:2018bef}. 
We will illustrate these ideas with specific calculations shortly in the next section.
 
 \section{Entanglement entropy from field theory}\label{sec3}

Now with all ingredients in place,  we proceed to compute the entanglement entropy from the field theory side using the formula  (\ref{SEE}). To do that, as explained earlier, we need to first compute the sphere partition function of the field theory on a sphere of arbitrary $R.$
Substituting  (\ref{Xdef}) in (\ref{eq2}) one can solve for the $\omega_d(R)$. This will lead us to a quadratic equation in $\omega_d(R)$ leading to two solutions.  
Following \cite{Donnelly:2018bef},  we will choose the negative sign in the solution as that will reproduce the known CFT  results in $\lambda_d \rightarrow 0$ limit (in even dimension it will give the well known CFT anomaly results). At the end of this section we have made some further comment on the other branch of solution, namely the positive branch of $\omega_d(R)$. 

It is clear now that solving the differential equation (\ref{eq2.9}) with the negative branch of $\omega_d(R)$ one can calculate $\log Z_{S^d}$. To solve the differential equation \eqref{eq2.9} we have used the boundary condition $\log Z_{S^d}=0$ at $R=0$ \cite{Donnelly:2018bef}. Once we know $\log Z_{S^d}$, equation  (\ref{SEE}) enables us to compute exact entropy $S_{d,EE}$. Below we quote the results for computation in various dimensions.  Again, for the expressions of $\log Z_{S^d}$ and consequently of $S_{d,EE}$ for general $d$ please refer to the Appendix \ref{appendix}.

\subsection*{Case 1: $d=2$}
As an warmup exercise, let us first derive the results of \cite{Donnelly:2018bef} in our convention. Equation \eqref{eq2} and \eqref{X2} gives
\be{}
\omega_2(R)= \frac{1}{4\,\lambda_2}\Big(1-\sqrt{1+\frac{c \lambda_2 }{3 \pi  R^2}}\Big).
\ee
Here the central charge $c$ is related to the bulk quantities via (\ref{2c}). The sphere partition function in this case takes the form,
\be{}
\log~Z_{S^2} =\frac{c}{3} \, \sinh ^{-1}\left(\sqrt{\frac{3 \pi  R^2}{c\, \lambda_2 }}\right)+ \frac{R^2}{3\,\lambda_2}\Big( \sqrt{9 \pi ^2+\frac{3 \pi  c\, \lambda_2 }{R^2}}-3 \pi\Big).
\ee
Then from (\ref{SEE}), we can operate the differential operator on the above equation to get the entanglement entropy for the desired region,
\be{}
S_{2,EE}=\frac{c}{3}  \sinh ^{-1}\left( \sqrt{\frac{3 \pi\, R^2  }{c\, \lambda_2 }}\right).
\ee

A pertinent check is to recover the CFT$_2$ result back from this expression. As we explained earlier, our expansion parameter  is  $t_2=c\lambda_2 $. On expanding the above result in the limit $R\gg
\sqrt{t_2}$ we get the well known result of entanglement entropy of a CFT$_2$:
\be{}
S_{2,EE}  = \frac{c}{3}\log\left(\sqrt{\frac{12\pi}{t_2}}R  \right) + \frac{c^2 \lambda _2}{36 \pi  R^2} + \mathcal{O}(t_2^2).
\ee
Simply following the two dimensional case, we will see that under expansion with right parameters, we will always get back the known CFT$_d$ results in all $d\geq 2$ \cite{Cardy1} as well.
\subsection*{Case 2: $d=3$}
In three dimensions,  the deformation operator takes the form,
\begin{eqnarray}
X_3= \left(T_{ab}+\frac{\alpha_3}{\lambda_3^{1/3}}\mathcal{G}_{ab}\right)^2-\frac{1}{2}\left(T^a_a+\frac{\alpha_3}{\lambda_3^{1/3}}\mathcal{G}^a_a\right)^2.
\end{eqnarray}
Plugging this in \eqref{eq2} results in getting one of the solutions as,
\be{}
\omega_3(R)=\frac{R^2+3\,t_3- R \sqrt{R^2+6\, t_3}}{3\, \lambda_3\,  R^2},
\ee
where we use the parameter $t_3=\alpha_3\lambda_3^{2/3}$. 
 
The sphere partition function takes the form
\be{}
\log~Z_{S^3} = \frac{2 \pi ^2 \left(-R^3+R^2 \sqrt{R^2+6 t_3}+6 t_3 \sqrt{R^2+6 t_3}-9 R t_3\right)}{3 \lambda_3 }.
\ee
Quite straightforwardly, the entanglement entropy in this case is given by
\be{3eeft}
S_{3,EE}=\frac{4 \pi ^2 t_3 \left(R^2+6 \,t_3-R \sqrt{R^2+6\, t_3}-\sqrt{6\, t_3 \left(R^2+6\, t_3\right)}\right)}{\lambda_3  \sqrt{R^2+6\, t_3}}.
\ee
Expanding the entanglement entropy for $R\gg \sqrt{t_3}$ one recovers the results for a CFT$_3$ entanglement:
\be{3,EEx}
S_{3,EE}=-\frac{4 \sqrt{6} \pi ^2 t_3^{3/2}}{\lambda_3 }+\frac{12 \pi ^2 t_3^2}{\lambda_3  R}+\mathcal{O}(t_3^{5/2}).
\ee
As expected there are no logarithmic terms in the expression \cite{KL,KL1,KL2}.  We will come back to the expression of entanglement entropy in three dimensions, in the next section, where we compare the field theory results to those obtained from holography.

\subsection*{Case 3: $d=4$}
We move on to the most interesting case: four dimensional field theory. The deformation operator is given by,
\begin{eqnarray}
X_4= \left(T_{ab}+\frac{\alpha_4}{\lambda_4^{1/2}}\mathcal{G}_{ab}\right)^2-\frac{1}{3}\left(T^a_a+\frac{\alpha_4}{\lambda_4^{1/2}}\mathcal{G}^a_a\right)^2.
\end{eqnarray}
Solving \eqref{eq2} with the $X_4$ from the above equation gives
\be{}
 \omega_4(R)= \frac{3 \left(R^2+8\, t_4 - R \sqrt{R^2+16 t_4}\right)}{8 \,\lambda_4\,  R^2},
 \ee
 where $t_4 = \alpha_4 \sqrt{\lambda_4}$.
The sphere partition function can be calculated in the following form
 \be{}
\log~Z_{S^4} =\frac{\pi ^2 \left(R \left(-R^3+R^2 \sqrt{R^2+16 t_4}+8 t_4 \sqrt{R^2+16 t_4}-16 R t\right)-128\, t_4^2\, \sinh ^{-1}\left(\frac{R}{4 \sqrt{t_4}}\right)\right)}{\lambda_4 }.
\ee
Following usual procedure, the field theory entanglement entropy takes the following form:
 \be{4,EE}
S_{4,EE}= -\frac{8 \pi ^2 t_4 \left(R \left(R-\sqrt{R^2+16 t_4}\right)+16 \, t_4 \sinh ^{-1}\left(\frac{R}{4 \sqrt{t_4}}  \right)\right)}{\lambda_4 }.
\ee
To arrive at CFT$_{4}$ results, we expand $S_{4,EE}$ small $t_4$, which spells out,
\be{4,EEx}
S_{4, EE}= -\frac{64 \pi ^2 t_4^2}{\lambda_4 } \left[2\log \left(\frac{R}{2\sqrt{t_4}}\right)-1\right]+\mathcal{O}(t_4^3).
\ee
Once again we recover the universal logarithmic term in the entanglement entropy of a CFT$_{4}$ with the exact coefficients attached.
\subsection*{Case 4: $d=5$}
The deformation operator in five dimensions, as discussed earlier, starts having contributions from the $\mathcal{C}_{ab}$ tensor, leading to the form,
\begin{eqnarray}
X_5= \left(T_{ab}+\frac{\alpha_5}{\lambda_5^{3/5}}\mathcal{C}_{ab}\right)^2-\frac{1}{4}\left(T^a_a+\frac{\alpha_5}{\lambda_5^{3/5}}\mathcal{C}^a_a\right)^2+\frac{\alpha_5}{5\lambda^{8/5}}\left(\frac{3}{2}\mathcal{R}+\mathcal{C}^a_a\right).
\end{eqnarray}
Usual procedure to calculate $\omega_5(R)$ yields
\be{}
\omega_5(R)= \frac{2 R^4+30 R^2 t_5 -225 t_5^2- 2 R^3 \sqrt{R^2+30 t_5}}{5\, \lambda_5\,  R^4}.
\ee
Using this expression one can calculate the sphere partition function, and hence the entanglement entropy. The expressions are given respectively by
\begin{eqnarray}
\log~Z_{S^5} &=& -\frac{2\pi^3\, R^3}{5\lambda}\Big[ R^2-R\sqrt{R^2+30\,t_5}+25\, t_5-\frac{10 \, t_5\, \sqrt{R^2+30\,t_5}}{R}\\ \nonumber &&-\frac{1125\, t_5^2}{2\, R^2} +\frac{600\,t_5^2}{R^3}\Big(\sqrt{30\,t_5}-\sqrt{R^2+30\,t_5}\Big)\Big] , \\ 
S_{5, EE}&=& \frac{4 \pi ^3 t_5}{\lambda_5  \sqrt{R^2+30 t_5}} \Big[R^4+60 \left(\sqrt{30} \sqrt{t_5^3 \left(R^2+30 t_5\right)}-30 t_5^2\right)\\ \nonumber
&-&30 R^2 t_5+45 R t_5 \sqrt{R^2+30 t_5}-R^3 \sqrt{R^2+30 t_5}\Big].
\end{eqnarray}
We can go to the CFT limit by expanding in $t_5$ to get,
\be{}
S_{5,EE} = \frac{240 \sqrt{30} \pi ^3 t_5^{5/2}}{\lambda_5 }-\frac{4050 \pi ^3 t_5^3}{\lambda_5  R}+\mathcal{O}(t_5^4).
\ee
\subsection*{Case 4: $d=6$}
The six dimensional deformation operator here takes the form,
\begin{eqnarray}
X_6= \left(T_{ab}+\frac{\alpha_6}{\lambda_6^{2/3}}\mathcal{C}_{ab}\right)^2-\frac{1}{5}\left(T^a_a+\frac{\alpha_6}{\lambda_6^{2/3}}\mathcal{C}^a_a\right)^2+\frac{\alpha_6}{6\lambda_6^{5/3}}\left(2\mathcal{R}+\mathcal{C}^a_a\right).
\end{eqnarray}
As in the other cases, the $\omega_6(R)$ is calculated as,
\be{}
\omega_6(R)=\frac{5 \left(R^4+24 R^2\,  t_6-288 t_6^2 - R^3 \sqrt{R^2+48 t_6}\right)}{12\, \lambda_6\,  R^4}.
\ee
The sphere partition function reads as,
\bea{}\nonumber
\log~Z_{S^6} =\frac{4 \pi ^3}{9 \lambda_6 } \Bigg[-R^6&-&36 R^4 t_6 + 864 R^2 t_6^2+R \left(R^2+36 t_6\right) \sqrt{R^2+48 t_6} \left(R^2-24 t_6\right)\\  &+& 41472 t_6^3 \sinh ^{-1}\left(\frac{R}{4 \sqrt{3} \sqrt{t_6}}\right)\Bigg] .
\eea
And, again we arrive at the expression for the entanglement entropy by operating with the right differential operator,
\bea{}
S_{6,EE}=\frac{16 \pi ^3 t_6}{3 \lambda_6 }
 \Bigg[R \left(-R^3+R^2 \sqrt{R^2+48\, t_6}-72 t_6 \sqrt{R^2+48 t_6}+48 R\, t_6\right)\\ \nonumber + 3456\, t_6^2 \sinh ^{-1}\left(\frac{R}{4 \sqrt{3} \sqrt{t_6}}\right)\Bigg].
 \eea
Expanding for small  $t_6$ we recover the CFT result;
\be{}
S_{6,EE}=\frac{1536 \pi ^3 t_6^3}{\lambda_6 } \left[6 \log \left(\frac{R}{2\sqrt{3}\, \sqrt{t_6}}\right)-7\right]+\mathcal{O}(t_6^4).
\ee
Now once we have quoted all the results obtained in this section, some important comments are in order. Let us discuss them one by one as follows,
\begin{itemize}
\item{The exact entanglement entropy calculated in this section does not contain non-universal terms, in particular when expanded around the CFT limit, for example in (\ref{3,EEx}), it is free of power law terms (scaling with positive powers of $R$). Note that these results do not capture the well known \textit{`area law'} as the counter terms cancels those out. The polynomial terms (for $d\geq 3$) in entanglement entropy are not universal and depends on the choice of the renormalization scheme. So, by appropriate choice of counter terms, one can get rid of these non-universal pieces in the entanglement entropy.  This analysis is studied in considerable detail in \cite{RG,RG1, RG2}. In our calculation, the trace of the stress tensor \eqref{Xdef} assumes a particular choice of renormalization scheme so that only finite terms show up in the answer. This is reflected in our result of finite partition function and hence works similarly for the entanglement entropy.

To recover non-universal terms in the entanglement entropy in our setup, one needs to do away with the counter terms $S_{ct}$ in the bulk action. This will bring back scheme dependent 
power law terms in the entanglement entropy i.e. area terms will pop up again in analogues of (\ref{3,EEx}).
We can turn off all these counter terms simply by putting $c_{i}=0$ in (\ref{ict}), see Appendix  \ref{appendix} for a detailed discussion.
 In the section next to the current one, we will perform holographic calculation of entanglement entropy which, not surprisingly, will contain non-universal terms along with the universal pieces computed from field theory. We will comment about this more in the next section.}
\item{Secondly, we would like to mention the fact that solving \eqref{eq2} gives two different solutions of $\omega_d(R)$. We choose the negative branch because in the limit $\lambda_d\to 0$ we recover the well known trace anomaly of a CFT in various dimensions. But, it is likely that the other branch, namely the positive branch is related to the non-perturbative states of the deformed theory. These states decouple from the original undeformed CFT in the limit $\lambda_d\to 0$. In the case of $d=2$, this has been investigated in the context of modular invariance of the partition function on a torus \cite{Aharony:2018bad} (also see \cite{Aharony:2018ics}). Possibly, this would also give rise to non-perturbative ambiguities in $d\geq 3$ as discussed in \cite{Aharony:2018bad,Aharony:2018ics} in the case of $d=2$.}
\end{itemize}

\section{Entanglement entropy: A holographic setup}\label{sec4}
In this section we compute the holographic entanglement entropy using the Ryu-Takayanagi prescription \cite{Ryu, Ryu1}. We are interested in calculating the entanglement entropy between two codimension-one hemispheres. The entangling surface in this setup would be a codimension-two sphere. From the holographic side one needs to compute the minimal area hypersurface in $AdS_{d+1}$ whose boundary coincides with the entangling surface $S^{d-1}$ residing on the cut-off surface of $AdS_{d+1}$.  To facilitate the computation we start with global $AdS_{d+1}$ with the metric
\begin{eqnarray}
ds^2=L^2\Big(d\rho^2+\sinh^2\rho(d \theta_1^2+\sum_{j=2}^{d} \prod_{i=1}^{j-1}\sin^2 \theta_i d \theta_j^2)\Big).
\end{eqnarray}
Then using the RT formula we get the desired entanglement entropy as,
\begin{align}
\begin{split} \label{compare}
S_{d, EE} & =\frac{2\,\pi L^{d-1} \Omega_{d-2}}{\lp^{d-1}}\int_0^{\rho_c} \sinh^{d-2}(\rho) d\rho \\
&=\frac{2\pi\, L^{d-1} \Omega_{d-2} }{\lp^{d-1}} ~
\frac{\sinh^{d-1}(\rho_c)}{d-1}\,\, {}_2F_1\left[\frac{1}{2},\frac{d-1}{2},\frac{d+1}{2};-\sinh^2(\rho_c)\right]
 \end{split}
\end{align}
where $\Omega_{d-2}=\frac{2\pi^{(d-1)/2}}{\Gamma(\frac{d-1}{2})}$ is the area of a unit $d-2$ dimensional sphere $S^{d-2}$ and $\sinh (\rho_c)=\frac{R}{L},$ where $R$ is the radius of entangling region. $L$ is the AdS radius and $\lp$ is the Planck length.
We now can explicitly write the total result for the entropy in various dimensions from the expressions of $S_{d.EE}$. 
\begin{align}
\begin{split}
& d=2: S_{2,EE}=\frac{4\,\pi\, L \sinh ^{-1}\left(\frac{R}{L}\right)}{\lp},\\&
d=3: S_{3,EE}=\frac{4 \pi ^2 L \left(\sqrt{R^2+L^2}-L\right)}{\lp^2},\\&
d=4: S_{4,EE}=\frac{4 \pi ^2 L \left(R \sqrt{R^2+L^2}-L^2 \sinh ^{-1}\left(\frac{R}{L}\right)\right)}{\lp^3},\\&
d=5:  S_{5,EE}=\frac{4 \pi ^3 L \left(2 L^3-2 L^2 \sqrt{L^2+R^2}+R^2 \sqrt{L^2+R^2}\right)}{3 \lp^4},\\&
d=6: S_{6,EE}= \frac{2 \pi ^3 L \left(R \sqrt{L^2+R^2} \left(2 R^2-3 L^2\right)+3 L^4 \sinh ^{-1}\left(\frac{R}{L}\right)\right)}{3 \lp^4}.
\end{split}
\end{align}

Now the test here is to compare the holographic results explicitly with the field theory result for all these cases. Take for example the case of $d=2$. Given the dictionary $\frac{4\,\pi\, L }{\ell_P} = \frac{c}{3}$ and $c\lambda_2 = 3\pi L^2$, one finds an exact match from both sides. This has already been established in \cite{Donnelly:2018bef} even at subleading level.

In $d=3$, things turn out to be more interesting. The dictionary here can be constructed from (\ref{relation}), which reads $L= \sqrt{6\, t_3}$ and $\lambda_3 =\frac{1}{6} \ell_P^2 L$. Using the dictionary, the holographic result expressed in terms of field theory variables takes the form,
\be{}
S_{3,EE} = -\frac{4 \sqrt{6} \pi ^2 t_3^{3/2}}{\lambda_3 }+\frac{4 \pi ^2 t_3 \sqrt{R^2+6 t_3}}{\lambda_3 }.
\ee
Comparing the above result with (\ref{3eeft}), we can observe while other terms match perfectly, it seems as if result from field theory captures an ``extra'' term $\frac{-4\pi^2 t_3 R}{\lambda_3}$. This particular term looks like the $area~term$ in three dimensions, albeit with an opposite sign. Obviously, some comments are in order here on the role of this ``extra'' term. 

Let us explain a bit more about this area term and universal pieces in entanglement entropy for the undeformed theory for the convenience of the reader. If we can assume the presence of a UV cut-off $r_c$ as some finite radial cut-off, then the leading contribution in entropy can always be written as \cite{SM},
\be{SM}
S_{d,EE}\approx \frac{2\pi}{\pi^{d/2}}\frac{\Gamma(d/2)}{d-2}a_d \frac{\mathcal{A}_{d-2}}{r_c^{d-2}}+...,
\ee
where $\mathcal{A}_{d-2} = \Omega_{d-2}R^{d-2}$  is the `area' of the equator dividing the two halves of the  $(d-1)$  dimensional sphere placed at the cut-off surface. The form of the universal contribution to the entanglement entropy, on the other hand, depends on whether we are working in odd or even dimension. Quoting well known results, the structure of the universal terms can be written as,
\bea{}
S_{univ}&=& (-1)^{\frac{d}{2}-1}~4a_d~\log\left(\frac{R}{r_c}\right),~~~~\text{For even dimensions}\\ \nonumber
&=& (-1)^{\frac{d-1}{2}}~2\pi a_d,~~~~~~~~~~~~~~~~~~\text{For odd dimensions}.
\eea
Here $a_d$ is the parameter identified in \cite{SM, me1} that characterizes monotonicity in holographic RG flows and in our notations has a form,
\be{adef}
a_d = \frac{\pi ^{d/2} L^{d-1}}{\lp^{d-1} \Gamma \left(\frac{d}{2}\right)}.
\ee
These universal terms are certainly the most interesting ones to be calculated in entanglement entropy computations.

 Now coming back to our three dimensional case, and using the proper dictionary, we can see that the ``extra'' term $\frac{-4\pi^2 t_3 R}{\lambda_3}$ is exactly the RHS of (\ref{SM}) in three dimensions, but with a negative sign. This term has a very unique effect, when the 
field theory result in $d\geq 2$ is expanded in the CFT limit this negative term explicitly cancels out the right-sign area term contribution, and thus in (\ref{3,EEx}) we see there is no term scaling with positive powers of $R$.

Let us also consider the case of $d=4$ to illustrate this phenomenon further. In this case, using the appropriate holographic dictionary, one can check that the entropy calculated from RT formula in terms of field theory variables becomes,
\be{}
S_{4,EE} = \frac{8 \pi ^2 R\, t_4 \sqrt{R^2+16 t_4}}{\lambda_4 }-\frac{128 \pi ^2 t_4^2 \sinh ^{-1}\left(\frac{R}{4 \sqrt{t_4}}\right)}{\lambda_4 }.
\ee
One can again compare this with the exact field theory result (\ref{4,EE}) and see the terms match perfectly with all prefactors in place. However, there is again one additional term in field theory computation, having a form $-\frac{8\pi^2R^2 t_4}{\lambda_4}$, which is explicitly the area term in four dimensions as explained before \footnote{One has to reinstate the factors of $r_c$ as usual here.}, again with the opposite sign. This term again cancels out the right-sign area term in the CFT limit expansion (\ref{4,EEx}). 
One can check that the trends persist in all other dimensions too, i.e. the universal terms in both field theory and RT calculation gives \textit{exact agreement} under the dictionary we propose, albeit with the extra area-like term which appears with a negative sign.\par

We must reiterate here, this apparent ``mismatch'' is merely an artefact of the renormalization scheme chosen to calculate the partition function, or in other words the chosen form of counter term action. The addition of the counter terms make sure there is no polynomial terms in the field theory answer, while RT formula naturally contains such power law terms. If we simply did not add any counter terms, then obviously both field theory and RT calculation would have these area terms in the expansion around the undeformed theory. Moreover, in Appendix \ref{appendix}, we clearly show how the results from both sides in our setup match exactly at the functional level if all counter terms are simply turned off. We also should mention that the case of two dimensions is a special one in this aspect. Although in \cite{Donnelly:2018bef}  the authors do not use any counter terms in their computation, the renormalization scheme we have used in (\ref{ict}) allows for one counter term to be present in the gravity action i.e. the one with coefficient $c_1$. One can show that the special structure of this counter term generates a piece with positive power of $R$ in $
\log Z$ (see Appendix \ref{appendix} for details), which in turn gets identically cancelled by the action of $(1-\frac{R}{d}\frac{d}{dR})$ operator. Hence, the presence of renormalization terms in 2d doesn't even matter for EE computation.

 Despite all these subtleties, it is always more appropriate to compute an observable analogue to entanglement entropy without these divergent pieces while showing a match from both sides of the duality. This is exactly what we will do in the next section.

%


\section{Comments on renormalized entanglement entropy}\label{sec5}

From our experiences with entanglement entropy in last two sections for the deformed theory, one would like to find an observable on both sides of the duality without any subtlety, and there seems to be one such observable that does the job for us.
Following \cite{LiuMZ, LiuMZ1} we compute the Renormalized Entanglement Entropy ($S_{d, REE}$) both from the field theory and holographic side.  REE is defined as follows,
\begin{align}
\begin{split} \label{REE}
&S_{d,REE}=\frac{1}{(d-2)!} R\frac{d}{dR} (R\frac{d}{dR}-2)\cdots(R \frac{d}{dR}-(d-2))~S_{d,EE} \quad ~~~~~~~\rm{for\, even\,} d,\nonumber \\&
S_{d,REE}=\frac{1}{(d-2)!}(R\frac{d}{dR}-1)(R\frac{d}{dR}-3)\cdots(R \frac{d}{dR}-(d-2))~S_{d,EE} \quad \rm{for\, odd \,} d.
\end{split}
\end{align}
Advantage of this quantity as described in \cite{LiuMZ, LiuMZ1} that it efficiently picks out the universal  pieces in the entanglement entropy. So for our case it will automatically be free of the area terms (both leading and subleading), which is important following our discussion in the last section. Also it removes all the IR divergences (if any) from the computed quantity, however that will not be important for our case.  As we have seen from the previous sections that the way we have formulated the field theory computation, the $S_{d,EE}$ coming from it does not include the area terms, while the computations from RT formula indeed does. Hence  REE seems to be a reasonable quantity which we can use to  make comparison between holography and field theory results.  Below we quote the result for REE computed  in different dimensions using the above definition.
\begin{align}
\begin{split}
& d=2: S_{2,REE} =\frac{\sqrt{\pi } c\, R}{\sqrt{3\, c\, \lambda_2 +9 \pi  R^2}} ,\\&
d=3: S_{3,REE}=\frac{4 \pi ^2 t_3^{3/2} \left(\sqrt{6}-6 \sqrt{\frac{t_3}{R^2+6 t_3}}\right)}{\lambda_3 },\\&
d=4: S_{4,REE}= \frac{128\, \pi ^2 R^3\, t_4^2}{\lambda_4  \left(R^2+16 \,t_4\right)^{3/2}},\\&
d=5: S_{5,REE}=\frac{120 \pi ^3 t_5^{3/2} \left(-900 \sqrt{t_5^5 \left(R^2+30 t_5\right)}-45 R^2 \sqrt{t_5^3 \left(R^2+30 t_5\right)}+\sqrt{30}\, t_5 \left(R^2+30 t_5\right)^2\right)}{\lambda_5  \left(R^2+30 t_5\right)^2},\\&
d=6: S_{6,REE}=\frac{6144 \, \pi ^3 R^5\, t_6^3}{\lambda_6  \left(R^2+48 \, t_6\right)^{5/2}}.
\end{split}
\end{align}
Again to emphasize, we have computed REE for both the field theory and holography using the results of the  Section~(\ref{sec3}) and Section~(\ref{sec4}). We find \textit{prefect agreement} between the expressions in all the cases. This also serves as a nice consistency check of our computations. 

Now from the expressions written above, we can easily see in the UV limit ($R\rightarrow 0$) \footnote{We do not mention here the constant of proportionality without any loss of generality. },
\begin{align}
\begin{split} \label{eq5}
&S_{d, REE}=0.
\end{split}
\end{align}
And similarly, in the IR limit $(R\to\infty)$,
\begin{align}
\begin{split} \label{eq6}
S_{d,REE} \approx &\, a_d+\mathcal{O}\left(\frac{1}{R^2}\right), \quad \rm{even}\, dimensions,\\
\approx & ~a_d+\mathcal{O}\left(\frac{1}{R}\right), \quad \rm ~{odd}\, dimensions.  
\end{split}
\end{align}
where $a_d$ is defined in (\ref{adef}) in terms of gravitational parameters. For even dimensions, it  coincides with coefficients of the Euler anomaly terms of CFT at IR \cite{Solo1, Solo2, Ryu, Ryu1, SM,SM1}. For odd dimensions it is equal to the finite part of the free energy on sphere (independent of $R$ and cut-off) \cite{KL, KL1,KL2,SM, SM1}. Note that originally (in \cite{LiuMZ, LiuMZ1}) REE is defined for local field theories with two fixed points. But here we extend this idea to a situation where the high energy behaviour of the theory is not governed by a fixed point but flows to an IR- CFT fixed point. The energies of the highly excited states of the undeformed CFT becomes complex upon deformation \cite{McGough:2016lol,Taylor:2018xcy}. Hence we don't expect these states to add to the real degrees of freedom of the deformed theory. So the absence of any real degrees of freedom at high energy is being captured by the equation (\ref{eq5}) \footnote{However in the full theory with all the non-perturbative effects taken into account, it could be likely that our expressions for small values of $R$ will receive corrections. Commenting on the type of non-perturbative corrections is beyond the scope of this paper. For related discussion in $d=2$ see \cite{Aharony:2018bad}.}.  At low energy $(R\to\infty)$ it coincides with usual CFT result.  Also (\ref{eq5}) and (\ref{eq6}) are showing that the real degrees of freedom 
are increasing as the theory flows down the RG. This matches with our intuitive notion explained above. It will be an interesting future problem to investigate the exact nature of the flow in terms of entanglement entropy.

\section{Conclusions and discussions}
Let us start with a summary of the paper. In this work, inspired by recent developments in integrable deformations of CFTs and relevant holographic setups, we have calculated the exact entanglement entropy in $d$-dimensional CFTs deformed by an irrelevant operator quadratic in
components of the stress tensor. This reaffirms and extends the work in \cite{Donnelly:2018bef} to higher dimensional cases, where we could find the exact entanglement entropy both from the field theory partition 
function on a $d$-sphere and dual bulk calculation of RT surfaces in $AdS_{d+1}$ with  a hard cutoff along the radial direction. Our calculation also solidifies the holographic dual proposal of \cite{Taylor:2018xcy,Hartman:2018tkw} to higher dimensional theories, where we build up definitions of the deformation operators following \cite{Taylor:2018xcy,Hartman:2018tkw}. Exactly as in the $2d$ case, we find the entanglement entropy is finite 
and the bulk-boundary computations can be compared under proper holographic dictionary. 

Our calculation is based upon the symmetries of the entangling region in higher dimensions and uses the simplest possible case to test the correspondence. To be precise, the method to generalize this to other entangling surfaces is not clear, even in the two dimensional case.  It would be fascinating to uncover this mystery as a future study. Specifically, it is well know that for even dimensions ($d > 2$) two different kind of anomaly terms appear in the trace of stress tensor. Universal term of the entanglement entropy for spherical entangling surface (for the CFT) picks out the Euler anomaly term. But when evaluated for the cylindrical surface the universal term it picks out the other anomaly term - Weyl anomaly \cite{Solo2}. So it will be interesting to see similar kind of effects for our case. 

The field theory techniques we have adapted in this paper, only extracts the universal terms in the field theory computation, added with an area like-term with a negative sign. We explicitly matched these universal terms with the result coming from the RT surface calculation under our proposed holographic dictionary in  $d=2,3,4,5,6$. We then go ahead to define the REE \cite{LiuMZ, LiuMZ1} for our case, and find that REE in all dimensions can be exactly matched both from field theory and holographic perspective, solidifying our proposal even further.

 As in the case of two dimensions, we can explicitly see the entanglement vanishes in the limit $R\to 0$ in all dimensions and coincide with usual CFT results at $R\rightarrow \infty.$ This matches nicely with our intuition about the observable.  Rather excitingly, this agreement between universal terms in field theory and holographic result in this case provides us with a playing ground for the Surface/State correspondence proposed in \cite{Miyaji} where the authors generalize the notion of  holographic duality making it insensitive to the existence of the boundary of spacetime. In this conjecture, it has been argued that any codimension-two convex surface inside $AdS_{d+1}$ spacetime can be associated with a quantum state and the entanglement entropy associated with this region (w.r.t to the rest of the spacetime) can be computed using RT formula. In that sense, our results can be thought of as one concrete example of this conjecture. However one certainly needs to be cautious here and more investigations are needed along these lines.

Going forward with this, one can extend the study in various directions, since the arena of $TT$ deformations in higher dimensions is much unexplored at this point. A straightforward generalization is that of calculating the conical entropy following \cite{Donnelly:2018bef} which eventually leads to the Renyi entropy, which can also provide an independent check of the results obtained in the current work. We hope to report about this soon \cite{WP}.

An obvious generalization of the particular deformation we investigated in this paper would be to understand the $TT$ deformation in a massive QFT at least perturbatively. Also, it will be interesting to understand the exact nature of renormalization group using the techniques of REE along the lines of  \cite{LiuMZ1, RG, RG1,RG2, RG3, RG3a, RG4,RG5,RG6}.

One could in principle consider the effect of $TT$ deformation on the transport coefficients. An obvious avenue to proceed would be to study the changes  in KSS bound \cite{Kovtun} on viscosity over entropy density. It would be interesting to check if the violation of the KSS bound in the presence of anisotropy could be restored upon turning on a $TT$ deformation.  

Throughout this paper we have explored the the deformation of a generic large-$N$ CFT by the $TT$ operator for small positive coupling $\lambda_d$. It would be nice to have a concrete understanding of the deformation for the other sign of $\lambda_d$. Also, we have only included  the negative branch of $\omega_d$ in our calculation. Inspired from the analysis of \cite{Aharony:2018bad} in $d=2$, it is tempting to conjecture that the positive branch of $\omega_d$ is related to the non-perturbative states in the deformed theory. It would be worth investigating whether the other branch of $\omega_d$ gives rise to something sensible and exciting at the same time. This may eventually lead to non-perturbative understanding of the deformed theory. Understanding of the deformed theory beyond the planer limit (at finite $N$) would be very exciting as well. \par

Last but not the least, our calculation is evidently exhibiting that in general dimensions in the high energy limit of the theory, it becomes devoid of any real degrees of freedom. The entanglement entropy is vanishing at high energy limit and grows towards the IR fixed point. This reminds us of the recent construction of quantum circuits in the context of circuit complexity \cite{ jm} where one starts with direct product state and gradually incorporates entanglement into it (in $2d$ this is similar to the cMERA  \cite{cmera} construction). It will be interesting to compute the complexity along the lines of \cite{meS} to see the effect of this deformation and connect it to the physics of RG flow \footnote{For a different viewpoint on complexity and cut-off geomtries one could see \cite{Akha} and follow-up works. }. We sincerely hope to come back and explore these avenues in the near future.

\section*{Acknowledgements}

We thank M. Smolkin and M. Goykhman for useful discussions. The work of SC is supported in part by a center of excellence supported by the Israel Science Foundation (grant number 2289/18). AB is supported by JSPS Grant-in-Aid for JSPS fellows (17F17023).
Aritra Banerjee (ArB) is supported in part by the Chinese Academy of Sciences (CAS) Hundred-Talent Program, by the Key Research
Program of Frontier Sciences, CAS, and by Project 11647601 supported by NSFC. ArB would like to thank the ICTP HECAP for warm hospitality during the course of this project.

\appendix

\section{Expression for $\log Z_{S^d}$ for arbitrary dimensions ($d\geq 2$)}\label{appendix}

Here we present the expression for $\log Z_{S^d}$ for arbitrary dimensions ($d\geq 2$). First we write the expression for the tensor $\mathcal{C}_{ab}$ for arbitrary dimensions.

\begin{eqnarray} \nonumber
\mathcal{C}_{ab}&=&c_2~\mathcal{G}_{ab}+c_3\frac{2\,d\,t_d}{d-4}\Bigg[ 2\left(\mathcal{R}_{acbd}\mathcal{R}^{cd}-\frac{1}{4}g_{ab}\mathcal{R}_{cd}\mathcal{R}^{cd}\right)-\frac{d}{2(d-1)}\left(\mathcal{R}\mathcal{R}_{ab}-\frac{1}{4}g_{ab}\mathcal{R}^2\right)\Bigg],\\
 \end{eqnarray}
 
Here we remind the reader the coefficients from (\ref{ict}), $c_1=1$ and non-vanishing only for $d \geq  2$, $c_2=1$ and non-vanishing only for $d \geq  3$ and $c_3=1$ and non-vanishing only for $ d \geq 5.$ We use the expansion parameter $t_d$ as defined in (\ref{expand}). Then using (\ref{eq2}) we can solve for the general value of $\omega_d(R)$ which we quote below,
\begin{eqnarray}
\omega_d(R)=\frac{d-1}{2d\, \lambda_d}\left[c_1+\frac{c_2~t_d\, d(d-2)}{R^2}\left(1-\frac{c_3}{c_2}~\frac{t_d\, d(d-2)}{2 R^2}\right)\pm \sqrt{1+\frac{2d(d-2)\,t_d}{R^2}}\right].
\end{eqnarray}
 Note that we can get two solutions here, and again only consider the one with negative sign. Then using (\ref{eq2.9}) we get, 
\begin{eqnarray}
\log Z_{S^d}  \label{imp}
&=&\frac{\Omega_d\,R^d}{2\,\lambda_d}\Bigg[ \frac{\sqrt{2d(d-2)\,t_d}}{R} ~_2F_1\left[-\frac{1}{2},\frac{d-1}{2},\frac{d+1}{2};-\frac{R^2}{2d(d-2)\,t_d}\right]-c_1~\,\frac{d-1}{d}\nonumber \\
&&-\frac{c_2~d(d-1)t_d}{R^2}+\frac{c_3~d^2(d-1)(d-2)^2\, t_d^2}{(d-4)R^4}\Bigg],
\end{eqnarray}
where  $\Omega_d$ is the area of  a $d$ dimensional unit sphere $S^d$ is given by
\begin{eqnarray}
\Omega_d=\frac{2\pi^{\frac{d+1}{2}}}{\Gamma\left(\frac{d+1}{2}\right)}.
\end{eqnarray}
And the $_2F_1$ is the usual Hypergeometric function.
Armed with this expression, and using (\ref{SEE}) we can compute the entanglement entropy. The result is the following,
\begin{align}
\begin{split}
S_{d,EE}=& \textstyle \frac{\Omega_d R^{d-1}}{2\,\lambda _d}\Big[\sqrt{2 (d-2) d \, t_d} \,\,_2F_1\left(-\frac{1}{2},\frac{d-1}{2}, \frac{d+1}{2};-\frac{R^2}{2 (d-2) d \, t_d}\right)-\frac{(d-1) \sqrt{2 (d-2) d\, t_d+R^2}}{d}\Big]\\&\textstyle+\frac{4 (d-1) (d-2)^2\,d\, R^{d-4} t_d^2}{(d-4) \,\lambda_d} \left(c_3-\frac{(d-4)\, R^2}{2\,t_d \, d\, (d-2)^2}  c_2\right).
\end{split}
\end{align}
We can further simplify this expressions using identities involving Hypergeometric functions. Finally this leads to,
\begin{align}
\begin{split}  \label{imp1}
S_{d,EE}=& \frac{\Omega_d\, R^{d-1}}{2\,d\,\lambda _d}\Bigg[\sqrt{2 (d-2) d \, t_d} \,\,_2F_1\left(\frac{1}{2},\frac{d-1}{2}, \frac{d+1}{2};-\frac{R^2}{2 (d-2) d \, t_d}\right)\Bigg]+\\&\frac{4 (d-1) (d-2)^2\,d\, R^{d-4} t_d^2}{(d-4) \,\lambda_d} \left(c_3-\frac{(d-4)\, R^2}{2\,t_d \, d\, (d-2)^2}  c_2\right).
\end{split}
\end{align}Again we can easily check that  in $R\rightarrow 0$ limit $S_{d,EE}$ vanishes, in any general dimension.\par

Also, if we don't use the counter terms and work only with the bare partition function, we get by setting $c_1=c_2=c_3=0$ in (\ref{imp1}),
\begin{align}
\begin{split} S_{d,EE}=& \frac{\Omega_d\, R^{d-1}}{2\,d\,\lambda _d}\Bigg[\sqrt{2 (d-2) d \, t_d} \,\,_2F_1\left(\frac{1}{2},\frac{d-1}{2}, \frac{d+1}{2};-\frac{R^2}{2 (d-2) d \, t_d}\right)\Bigg].
\end{split}
\end{align}
Then using the dictionary (\ref{relation}) we can transform the above into,
\begin{align}
\begin{split} S_{d,EE}=&\frac{2\,\pi R^{d-1} \Omega_{d-2}}{(d-1)\, \lp^{d-1}}\Bigg[\sqrt{2 (d-2) d \, t_d} \,\,_2F_1\left(\frac{1}{2},\frac{d-1}{2}, \frac{d+1}{2};-\frac{R^2}{L^2}\right)\Bigg].
\end{split}
\end{align}
This is precisely same as the holographic result (\ref{compare})  obtained using RT formula. 



\begingroup\begin{thebibliography}{10}
\bibitem{Smirnov:2016lqw} 
  F.~A.~Smirnov and A.~B.~Zamolodchikov,
  ``On space of integrable quantum field theories,''
  Nucl.\ Phys.\ B {\bf 915}, 363 (2017)
  [arXiv:1608.05499 [hep-th]].
  
	
	
	
	\bibitem{Cavaglia:2016oda} 
  A.~Cavaglia, S.~Negro, I.~M.~Szecsenyi and R.~Tateo,
  ``$T \bar{T}$-deformed 2D Quantum Field Theories,''
  JHEP {\bf 1610}, 112 (2016)
  [arXiv:1608.05534 [hep-th]].

\bibitem{McGough:2016lol} 
  L.~McGough, M.~Mezei and H.~Verlinde,
  ``Moving the CFT into the bulk with $ T\overline{T} $,''
  JHEP {\bf 1804}, 010 (2018)
  [arXiv:1611.03470 [hep-th]].
 
\bibitem{Shyam:2017znq} 
  V.~Shyam,
  ``Background independent holographic dual to $T\bar{T}$ deformed CFT with large central charge in 2 dimensions,''
  JHEP {\bf 1710}, 108 (2017)
  [arXiv:1707.08118 [hep-th]].
  
\bibitem{Giveon:2017nie} 
  A.~Giveon, N.~Itzhaki and D.~Kutasov,
  ``$ \mathrm{T}\overline{\mathrm{T}} $ and LST,''
  JHEP {\bf 1707}, 122 (2017)
  [arXiv:1701.05576 [hep-th]].
  
\bibitem{Giveon:2017myj} 
  A.~Giveon, N.~Itzhaki and D.~Kutasov,
  ``A solvable irrelevant deformation of AdS$_{3}$/CFT$_{2}$,''
  JHEP {\bf 1712}, 155 (2017)
  [arXiv:1707.05800 [hep-th]].
  
\bibitem{Asrat:2017tzd} 
  M.~Asrat, A.~Giveon, N.~Itzhaki and D.~Kutasov,
  ``Holography Beyond AdS,''
  Nucl.\ Phys.\ B {\bf 932}, 241 (2018)
  [arXiv:1711.02690 [hep-th]].

\bibitem{Chakraborty:2018kpr} 
  S.~Chakraborty, A.~Giveon, N.~Itzhaki and D.~Kutasov,
  ``Entanglement Beyond $\rm AdS$,''
  arXiv:1805.06286 [hep-th].

\bibitem{Chakraborty:2018aji} 
  S.~Chakraborty,
  ``Wilson loop in a $T\bar{T}$ like deformed $\rm{CFT}_2$,''
  Nucl.\ Phys.\ B {\bf 938}, 605 (2019)
  [arXiv:1809.01915 [hep-th]].
    
    
\bibitem{Giribet:2017imm} 
  G.~Giribet,
  ``$T\bar{T}$-deformations, AdS/CFT and correlation functions,''
  JHEP {\bf 1802}, 114 (2018)
  [arXiv:1711.02716 [hep-th]].
  
\bibitem{Dubovsky:2017cnj} 
  S.~Dubovsky, V.~Gorbenko and M.~Mirbabayi,
  JHEP {\bf 1709}, 136 (2017)
  doi:10.1007/JHEP09(2017)136
  [arXiv:1706.06604 [hep-th]].
  
\bibitem{Dubovsky:2018bmo} 
  S.~Dubovsky, V.~Gorbenko and G.~Hernndez-Chifflet,
  ``$ T\overline{T} $ partition function from topological gravity,''
  JHEP {\bf 1809}, 158 (2018)
  [arXiv:1805.07386 [hep-th]].
  
\bibitem{Kraus:2018xrn} 
  P.~Kraus, J.~Liu and D.~Marolf,
  ``Cutoff AdS$_3$ versus the $T\bar{T}$ deformation,''
  JHEP {\bf 1807}, 027 (2018)
  arXiv:1801.02714 [hep-th].
  
\bibitem{Cardy:2018sdv} 
  J.~Cardy,
  ``The $T\overline T$ deformation of quantum field theory as a stochastic process,''
  arXiv:1801.06895 [hep-th].
  
\bibitem{Cottrell:2018skz} 
  W.~Cottrell and A.~Hashimoto,
  ``Comments on $T \bar T$ double trace deformations and boundary conditions,''
  arXiv:1801.09708 [hep-th].
  
\bibitem{Aharony:2018vux} 
  O.~Aharony and T.~Vaknin,
  ``The TT* deformation at large central charge,''
    JHEP {\bf 1805}, 166 (2018)
  arXiv:1803.00100 [hep-th].
  
\bibitem{Dubovsky:2018dlk} 
  S.~Dubovsky,
  ``A Simple Worldsheet Black Hole,''
  JHEP {\bf 1807}, 011 (2018)
  arXiv:1803.00577 [hep-th].


\bibitem{Bonelli:2018kik} 
  G.~Bonelli, N.~Doroud and M.~Zhu,
  ``$T\bar T$-deformations in closed form,''
  JHEP {\bf 1806}, 149 (2018)
  arXiv:1804.10967 [hep-th].

\bibitem{Datta:2018thy} 
  S.~Datta and Y.~Jiang,
  ``$T\bar{T}$ deformed partition functions,''
  arXiv:1806.07426 [hep-th].
  
\bibitem{Donnelly:2018bef} 
  W.~Donnelly and V.~Shyam,
  Phys.\ Rev.\ Lett.\  {\bf 121} (2018) no.13,  131602
  [arXiv:1806.07444 [hep-th]].
  
\bibitem{Babaro:2018cmq} 
  J.~P.~Babaro, V.~F.~Foit, G.~Giribet and M.~Leoni,
  ``$T\bar{T}$ type deformation in the presence of a boundary,''
  arXiv:1806.10713 [hep-th].
  
\bibitem{Conti:2018jho} 
  R.~Conti, L.~Iannella, S.~Negro and R.~Tateo,
  ``Generalised Born-Infeld models, Lax operators and the $\textrm{T} \bar{\textrm{T}}$ perturbation,''
  arXiv:1806.11515 [hep-th].
  
\bibitem{Chen:2018eqk} 
  B.~Chen, L.~Chen and P.~x.~Hao,
  ``Entanglement Entropy in $T\overline{T}$-Deformed CFT,''
  arXiv:1807.08293 [hep-th].
  

\bibitem{Aharony:2018bad} 
  O.~Aharony, S.~Datta, A.~Giveon, Y.~Jiang and D.~Kutasov,
  ``Modular invariance and uniqueness of $T\bar{T}$ deformed CFT,''
  arXiv:1808.02492 [hep-th].
  
  \bibitem{Cardy:2018jho}
  J.~Cardy,
  ``$T\overline T$ deformations of non-Lorentz invariant field theories,''
  arXiv:1809.07849 [hep-th].
  
\bibitem{Jiang:2019tcq} 
  Y.~Jiang,
  ``Expectation value of $\mathrm{T}\overline{\mathrm{T}}$ operator in curved spacetimes,''
  arXiv:1903.07561 [hep-th].
  
\bibitem{Park:2018snf} 
  C.~Park,
  Int.\ J.\ Mod.\ Phys.\ A {\bf 33}, no. 36, 1850226 (2019)
  doi:10.1142/S0217751X18502263
  [arXiv:1812.00545 [hep-th]].

\bibitem{Sun:2019ijq} 
  Y.~Sun and J.~R.~Sun,
  ``Note on Renyi entropy of 2D perturbed free fermions,''
  arXiv:1901.08796 [hep-th].

\bibitem{Wang:2018jva} 
  P.~Wang, H.~Wu and H.~Yang,
  ``The dual geometries of $T\bar{T}$ deformed CFT$_2$ and highly excited states of CFT$_2$,''
  arXiv:1811.07758 [hep-th].
  
  \bibitem{Gorbenko:2018oov} 
  V.~Gorbenko, E.~Silverstein and G.~Torroba,
  ``dS/dS and $ T\overline{T} $,''
  JHEP {\bf 1903}, 085 (2019)
  doi:10.1007/JHEP03(2019)085
  [arXiv:1811.07965 [hep-th]].

\bibitem{Araujo:2018rho} 
  T.~Araujo, E.~Colgain, Y.~Sakatani, M.~M.~Sheikh-Jabbari and H.~Yavartanoo,
  ``Holographic integration of $T \bar T $ \& $J \bar T$ via $O(d,d)$,''
  arXiv:1811.03050 [hep-th].

\bibitem{Guica:2017lia} 
  M.~Guica,
  ``An integrable Lorentz-breaking deformation of two-dimensional CFTs,''
  SciPost Phys.\  {\bf 5}, no. 5, 048 (2018)
  [arXiv:1710.08415 [hep-th]].

\bibitem{Bzowski:2018pcy} 
  A.~Bzowski and M.~Guica,
  ``The holographic interpretation of $J \bar T$-deformed CFTs,''
  JHEP {\bf 1901}, 198 (2019)
  [arXiv:1803.09753 [hep-th]].

\bibitem{Chakraborty:2018vja} 
  S.~Chakraborty, A.~Giveon and D.~Kutasov,
  ``$ J\overline{T} $ deformed CFT$_{2}$ and string theory,''
  JHEP {\bf 1810}, 057 (2018)
  [arXiv:1806.09667 [hep-th]].

\bibitem{Apolo:2018qpq} 
  L.~Apolo and W.~Song,
 ``Strings on warped AdS$_{3}$ via $ \mathrm{T}\bar{\mathrm{J}} $ deformations,''
  JHEP {\bf 1810}, 165 (2018)
  [arXiv:1806.10127 [hep-th]].

\bibitem{Nakayama:2018ujt} 
  Y.~Nakayama,
  ``Very Special $T\bar{J}$ deformed CFT,''
  arXiv:1811.02173 [hep-th].


\bibitem{Aharony:2018ics} 
  O.~Aharony, S.~Datta, A.~Giveon, Y.~Jiang and D.~Kutasov,
  ``Modular covariance and uniqueness of $J\bar{T}$ deformed CFTs,''
  JHEP {\bf 1901}, 085 (2019)
  [arXiv:1808.08978 [hep-th]].


\bibitem{Guica:2019vnb} 
  M.~Guica,
  ``On correlation functions in $J\bar T$-deformed CFTs,''
  arXiv:1902.01434 [hep-th].
  
\bibitem{Giveon:2019fgr} 
  A.~Giveon,
  ``Comments on $T\bar T$, $J\bar{T}$ and String Theory,''
  arXiv:1903.06883 [hep-th].
  
\bibitem{LeFloch:2019rut} 
  B.~Le Floch and M.~Mezei,
  ``Solving a family of $T\bar{T}$-like theories,''
  arXiv:1903.07606 [hep-th].

 


\bibitem{Taylor:2018xcy} 
  M.~Taylor,
  ``TT deformations in general dimensions,''
  arXiv:1805.10287 [hep-th].

\bibitem{Hartman:2018tkw} 
  T.~Hartman, J.~Kruthoff, E.~Shaghoulian and A.~Tajdini,
  ``Holography at finite cutoff with a $T^2$ deformation,''
  arXiv:1807.11401 [hep-th].

\bibitem{Caputa:2019pam} 
  P.~Caputa, S.~Datta and V.~Shyam,
  ``Sphere partition functions and cut-off AdS,''
  arXiv:1902.10893 [hep-th].

\bibitem{Zamolodchikov:2004ce} 
  A.~B.~Zamolodchikov,
  ``Expectation value of composite field T anti-T in two-dimensional quantum field theory,''
  hep-th/0401146.


\bibitem{Duff:1987cs} 
  M.~J.~Duff, T.~Inami, C.~N.~Pope, E.~Sezgin and K.~S.~Stelle,
  ``Semiclassical Quantization of the Supermembrane,''
  Nucl.\ Phys.\ B {\bf 297}, 515 (1988).

\bibitem{Ryu}
  S.~Ryu and T.~Takayanagi,
  ``Holographic derivation of entanglement entropy from AdS/CFT,''
  Phys.\ Rev.\ Lett.\  {\bf 96} (2006) 181602
  [hep-th/0603001].
	
\bibitem{Ryu1}
  S.~Ryu and T.~Takayanagi,
  ``Aspects of Holographic Entanglement Entropy,''
  JHEP {\bf 0608} (2006) 045
  [hep-th/0605073].

  \bibitem{Solo1}
   S.~N.~Solodukhin,
  ``Entanglement entropy of black holes and AdS/CFT correspondence,''
  Phys.\ Rev.\ Lett.\  {\bf 97} (2006) 201601
  [hep-th/0606205].
  
  \bibitem{Solo2}
 S.~N.~Solodukhin,
  ``Entanglement entropy, conformal invariance and extrinsic geometry,''
  Phys.\ Lett.\ B {\bf 665} (2008) 305
  [arXiv:0802.3117 [hep-th]].
 
 
 \bibitem{SM}
  R.~C.~Myers and A.~Sinha,
  ``Holographic c-theorems in arbitrary dimensions,''
  JHEP {\bf 1101} (2011) 125
  [arXiv:1011.5819 [hep-th]].
  
 \bibitem{SM1}
  R.~C.~Myers and A.~Sinha,
  ``Seeing a c-theorem with holography,''
  Phys.\ Rev.\ D {\bf 82} (2010) 046006
  [arXiv:1006.1263 [hep-th]].
  
  \bibitem{LiuMZ} 
  H.~Liu and M.~Mezei,
  ``A Refinement of entanglement entropy and the number of degrees of freedom,''
  JHEP {\bf 1304}, 162 (2013)
  [arXiv:1202.2070 [hep-th]].
	
	\bibitem{LiuMZ1}
  H.~Liu and M.~Mezei,
  ``Probing renormalization group flows using entanglement entropy,''
  JHEP {\bf 1401} (2014) 098
  [arXiv:1309.6935 [hep-th]].

\bibitem{renvol} 
  G.~Anastasiou, I.~J.~Araya, C.~Arias and R.~Olea,
  ``Einstein-AdS action, renormalized volume/area and holographic Rényi entropies,''
  JHEP {\bf 1808}, 136 (2018)
  doi:10.1007/JHEP08(2018)136
  [arXiv:1806.10708 [hep-th]].


\bibitem{Emparan:1999pm}
  R.~Emparan, C.~V.~Johnson and R.~C.~Myers,
 ``Surface terms as counterterms in the AdS / CFT correspondence,''
  Phys.\ Rev.\ D {\bf 60} (1999) 104001
  [hep-th/9903238].
 
  \bibitem{me}
  A.~Bhattacharyya, L.~Y.~Hung, K.~Sen and A.~Sinha,
  ``On c-theorems in arbitrary dimensions,''
  Phys.\ Rev.\ D {\bf 86} (2012) 106006
  [arXiv:1207.2333 [hep-th]].

 \bibitem{Balasubramanian:1999re}
  V.~Balasubramanian and P.~Kraus,
  ``A Stress tensor for Anti-de Sitter gravity,''
  Commun.\ Math.\ Phys.\  {\bf 208} (1999) 413
  [hep-th/9902121].

 
\bibitem{Central}
  J.~ D. Brown and M. ~Henneaux, 
  ``Central Charges in the Canonical Realization of
Asymptotic Symmetries: An Example from Three-Dimensional Gravity,"  Commun. Math.
Phys. 104 (1986) 207 -- 226.

 \bibitem{Cardy1}
  P.~Calabrese and J.~L.~Cardy,
  ``Entanglement entropy and quantum field theory,''
  J.\ Stat.\ Mech.\  {\bf 0406} (2004) P06002
  [hep-th/0405152].


\bibitem{KL}
  D.~L.~Jafferis, I.~R.~Klebanov, S.~S.~Pufu and B.~R.~Safdi,
  ``Towards the F-Theorem: N=2 Field Theories on the Three-Sphere,''
  JHEP {\bf 1106} (2011) 102
  [arXiv:1103.1181 [hep-th]].
  
  \bibitem{KL1}
   I.~R.~Klebanov, S.~S.~Pufu and B.~R.~Safdi,
  ``F-Theorem without Supersymmetry,''
  JHEP {\bf 1110} (2011) 038
  [arXiv:1105.4598 [hep-th]].
  
  \bibitem{KL2}
  I.~R.~Klebanov, S.~S.~Pufu, S.~Sachdev and B.~R.~Safdi,
  ``Entanglement Entropy of 3-d Conformal Gauge Theories with Many Flavors,''
  JHEP {\bf 1205} (2012) 036
  [arXiv:1112.5342 [hep-th]].



\bibitem{RG}
  T.~Faulkner,
  ``Bulk Emergence and the RG Flow of Entanglement Entropy,''
  JHEP {\bf 1505} (2015) 033
  [arXiv:1412.5648 [hep-th]]


\bibitem{RG1}
  O.~Ben-Ami, D.~Carmi and M.~Smolkin,
  ``Renormalization group flow of entanglement entropy on spheres,''
  JHEP {\bf 1508} (2015) 048
  [arXiv:1504.00913 [hep-th]].

\bibitem{RG2}
  C.~Akers, O.~Ben-Ami, V.~Rosenhaus, M.~Smolkin and S.~Yankielowicz,
  ``Entanglement and RG in the $O(N)$ vector model,''
  JHEP {\bf 1603} (2016) 002
  [arXiv:1512.00791 [hep-th]].


\bibitem{me1}
  A.~Bhattacharyya, M.~Sharma and A.~Sinha,
  ``On generalized gravitational entropy, squashed cones and holography,''
  JHEP {\bf 1401} (2014) 021
  [arXiv:1308.5748 [hep-th]].
  
  \bibitem{Miyaji}
  M.~Miyaji and T.~Takayanagi,
 ``Surface/State Correspondence as a Generalized Holography,''
  PTEP {\bf 2015} (2015) no.7,  073B03
  doi:10.1093/ptep/ptv089
  [arXiv:1503.03542 [hep-th]].

\bibitem{WP}
A.~ Banerjee, A.~Bhattacharyya and S.~Chakraborty, work in progress.

\bibitem{RG3}
  I.~R.~Klebanov, T.~Nishioka, S.~S.~Pufu and B.~R.~Safdi,
  ``Is Renormalized Entanglement Entropy Stationary at RG Fixed Points?,''
  JHEP {\bf 1210} (2012) 058
  [arXiv:1207.3360 [hep-th]].

\bibitem{RG3a}
  H.~Casini and M.~Huerta,
  ``On the RG running of the entanglement entropy of a circle,''
  Phys.\ Rev.\ D {\bf 85} (2012) 125016
  [arXiv:1202.5650 [hep-th]].

\bibitem{RG4}
  S.~Banerjee, Y.~Nakaguchi and T.~Nishioka,
  ``Renormalized Entanglement Entropy on Cylinder,''
  JHEP {\bf 1603} (2016) 048
  [arXiv:1508.00979 [hep-th]].

\bibitem{RG5}
  A.~Bhattacharyya, S.~Shajidul Haque and A.~Veliz-Osorio,
  ``Renormalized Entanglement Entropy for BPS Black Branes,''
  Phys.\ Rev.\ D {\bf 91} (2015) no.4,  045026
  [arXiv:1412.2568 [hep-th]].

\bibitem{RG6}
  M.~Taylor and W.~Woodhead,
  ``Renormalized entanglement entropy,''
  JHEP {\bf 1608} (2016) 165
  [arXiv:1604.06808 [hep-th]].

\bibitem{Kovtun}
   P.~Kovtun, D.~T.~Son and A.~O.~Starinets,
  ``Viscosity in strongly interacting quantum field theories from black hole physics,''
  Phys.\ Rev.\ Lett.\  {\bf 94} (2005) 111601
  [hep-th/0405231].

\bibitem{jm}
  R.~Jefferson and R.~C.~Myers,
  ``Circuit complexity in quantum field theory,''
  JHEP {\bf 1710} (2017) 107
  [arXiv:1707.08570 [hep-th]].

\bibitem{cmera}
  J.~Haegeman, T.~J.~Osborne, H.~Verschelde and F.~Verstraete,
  ``Entanglement Renormalization for Quantum Fields in Real Space,''
  Phys.\ Rev.\ Lett.\  {\bf 110} (2013) no.10,  100402
  [arXiv:1102.5524 [hep-th]].
  
  \bibitem{meS}
  A.~Bhattacharyya, A.~Shekar and A.~Sinha,
  ``Circuit complexity in interacting QFTs and RG flows,''
  JHEP {\bf 1810} (2018) 140
  [arXiv:1808.03105 [hep-th]].

\bibitem{Akha} 
  A.~Akhavan, M.~Alishahiha, A.~Naseh and H.~Zolfi,
  ``Complexity and Behind the Horizon Cut Off,''
  JHEP {\bf 1812}, 090 (2018)
  doi:10.1007/JHEP12(2018)090
  [arXiv:1810.12015 [hep-th]].
	
\end{thebibliography}\endgroup

\providecommand{\href}[2]{#2}

\end{document}